# Hyper-doping of Silicon for Plasmonics in the Telecommunication Range

*Jura Rensberg, Angela Barreda, Kevin Wolf, Andreas Undisz, Jürgen Salfeld, Sebastian Geburt, Isabelle Staude, Carsten Ronning, and Martin Hafermann**

J. Rensberg, A. Barreda, K. Wolf, I. Staude, C. Ronning, M. Hafermann
Friedrich Schiller University Jena, Institute of Solid-State Physics, Helmholtzweg 3, 07743 Jena, Germany
E-mail: martin.hafermann@uni-jena.de

A. Barreda, I. Staude
Friedrich Schiller University Jena, Institute of Applied Physics, Albert-Einstein-Straße 15, 07745 Jena, Germany

A. Barreda, I. Staude, C. Ronning
Friedrich Schiller University Jena, Abbe Center of Photonics, Albert-Einstein-Straße 6, 07745 Jena, Germany

A. Barreda
Group of Displays and Photonics Applications, Carlos III University of Madrid, Avda. de la Universidad, 30, Leganés, 28911 Madrid, Spain

A. Undisz
Chemnitz University of Technology, Institute of Materials Science and Engineering, Erfenschlager Straße 73, 09125 Chemnitz, Germany
and
Friedrich Schiller University Jena, Otto Schott Institute of Materials Research, Löbdergraben 32, 07743 Jena, Germany

J. Salfeld, S. Geburt
INNOVAVENT GmbH, Reinhard-Rube-Straße 4, 37077 Göttingen, Germany

Keywords: hyper-doping, plasmonics, telecommunication regime, pulsed laser melting, ion implantation

We investigate hyper-doping, a promising approach to introduce a high concentration of impurities into silicon beyond its solid solubility limit, for its potential applications in near-infrared plasmonics. We systematically explore the incorporation of dopants into silicon using ion implantation and pulsed laser melting annealing processes. Reflectance spectra analysis shows an achievable plasma wavelength of around 1.5 µm for dopant concentrations exceeding 4 at.%. Complex refractive index data for the doped silicon samples are extracted, revealing their potential for near-infrared plasmonic applications. Moreover, we propose a fabrication process that allows for the creation of hyper-doped silicon nanoparticles without the need for additional masking steps. Our research paves the way for designing CMOS-compatible plasmonic nanostructures operating in the telecommunication wavelength range. The study's findings offer significant insights into the utilization of hyper-doped silicon for advanced photonic and optoelectronic applications.

## 1. Introduction

Plasmonic nanoparticles have driven applications, including enhancing light-matter interactions, enabling ultrasensitive sensing, and manipulating light at the nanoscale. [1–5] Since conventional metallic components, such as noble metals, suffer from losses at optical frequencies, non-adjustable dielectric permittivity, and nanofabrication challenges, the search for more affordable and effective materials emerged. [6, 7] One promising candidate is n-type doped silicon (Si) as the cornerstone of semiconductor technologies, [8] whereas n-type doping is more suitable than p-type due to the smaller effective mass of conduction electrons compared to holes in Si. [9] Plasmonic behavior of n-type Si was observed mainly in the infrared spectral region for wavelengths >4 µm. [10–14] Lower plasma wavelengths, thus higher free carrier concentrations, are hindered by equilibrium processes and the solid solubility limit of Si. [15] Despite the ability to achieve doping concentrations larger than the solid solubility limit, [16] the free carrier concentrations are found to reach saturation at around $5 \times 10^{20}$ cm$^{-3}$. [17–20] Any additional introduced dopant atoms do not generate extra free carriers, due to the formation of dimers that induce localized deep-level states in the band gap, compensating the free carriers. To push the plasma wavelength even further into the NIR, active dopant concentrations must surpass the solid solubility limit, which is typically referred to as hyper-doping. This can be achieved only via non-equilibrium processes, such as rapid quenching of a melt or laser irradiation, leading to the formation of supersaturated solid solutions. The concept of hyper-doping has been experimentally demonstrated and has the potential to overcome limitations associated with conventional doping methods. [21–29] In addition, hyper-doping can result in significant changes in the material's electronic and optical properties, leading to new opportunities for applications in electronics, photonics, and optoelectronics. [23, 30] For example, hyper-doping of Si with deep impurities such as tellurium has been shown to result in the formation of a degenerated semi-metal band within the bandgap, which can be useful for solar cell and thermoelectric applications. [25, 31–33] To achieve hyper-doping, various techniques, such as ion implantation, molecular beam epitaxy (MBE), and solid-phase epitaxial recrystallization were used. [17, 34] These techniques are often combined with additional post-processing methods, such as low temperature annealing, to optimize the incorporation and activation of the dopants.

Here, we combine ion implantation of group V elements, i.e. phosphorous (P) and arsenic (As), and pulsed laser melting (PLM) annealing processes to achieve recrystallized hyper-doped n-type Si. Advantageously, using PLM alongside sub-µs rapid epitaxial recrystallization prevents dopant diffusion in the crystalline phase. We use a combination of Rutherford backscattering spectroscopy (RBS) in channeling mode, transmission electron microscopy (TEM), and optical reflectance spectroscopy to determine the maximal achievable amount of incorporated and electrically active dopants. By this means, we propose a fabrication process that allows for designing hyper-doped Si nanoparticles without the need for additional masking steps. To demonstrate the capabilities of our approach, we performed finite-difference time-domain (FDTD)-simulations to calculate the potential performance of such devices.

## 2. Results and Discussion

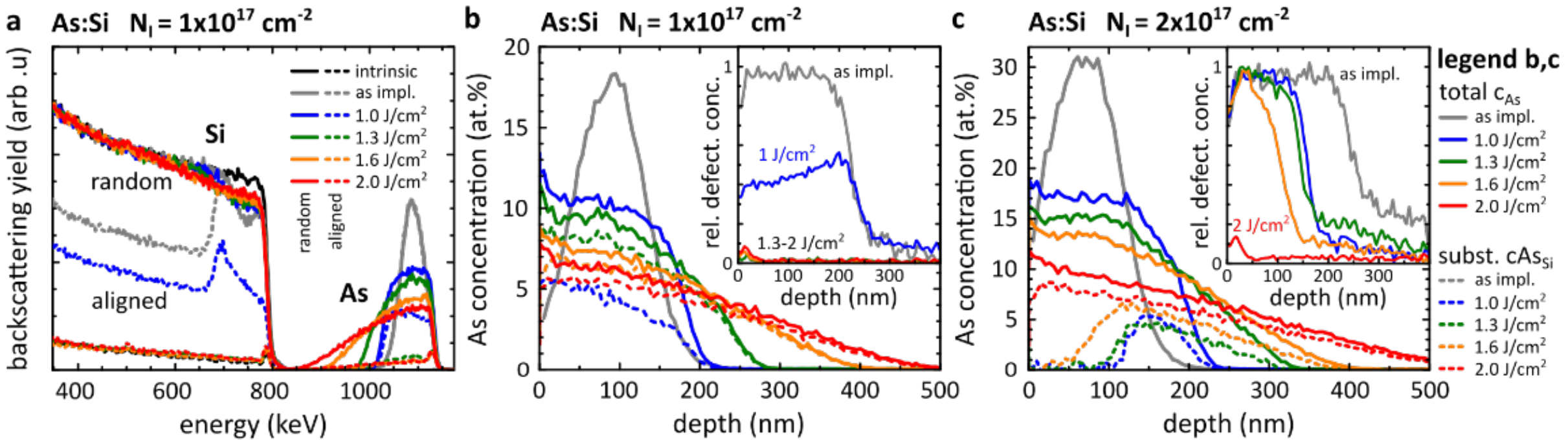

**Figure 1.** (a) Random and aligned RBS spectra of intrinsic Si compared to random and aligned RBS spectra of Si implanted with 150 keV $As^+$ to a fluence of $N_I = 1\text{x}10^{17}$ $cm^{-2}$ before (as implanted) and after subsequent pulse laser melting (PLM) annealing with different laser energy densities from 1.0 to 2.0 $J/cm^2$. (b) Concentration-depth profiles of As located in the Si matrix (solid lines, total amount $c_{As}$) and of As located on Si lattice sides (dashed lines, subst. $cAs_{Si}$) before and after PLM annealing extracted from the As signal of the RBS spectra given in (a), respectively. The corresponding relative defect concentrations within the Si matrix extracted from the Si part of the RBS spectra is shown in the inset of (b). (c) Same as (b) but for $N_I = 2\text{x}10^{17}$ $cm^{-2}$.

First, we systematically investigated the highest possible concentration of dopants that can be incorporated into the Si lattice with the combination of the non-equilibrium processes ion implantation and subsequent PLM annealing. **Figure 1**a exemplarily shows the RBS (random) and channeling spectra (aligned) measured for intrinsic Si, after $As^+$ implantation with $N_I$ = $1x10^{17}$ $cm^{-2}$, and after subsequent PLM annealing at various laser energy densities $E_L$ from 1.0 to 2.0 $J/cm^2$, respectively (see supplementary information Figure S1 for the RBS spectra of As:Si with $N_I$ = $2x10^{17}$ $cm^{-2}$). The As backscattering signal is well separated from the Si signal, which allows quantitative determination of the As depth distribution (Figure 1b and c). For this purpose, we used the ion beam analysis computer code Data Furnace, [35] which self-consistently extracts elemental depth profiles from RBS spectra. The total amount of As dopants located in the Si lattice as a function of depth *z* ($c_{As}$) was calculated from the random spectra. Whereas the amount of As dopants located on substitutional lattice sites ($cAs_{Si}$) was calculated from the difference between the random and corresponding aligned spectra. Dopants incorporated on regular lattice sites almost do not contribute to the backscattering yield of aligned spectra, whereas interstitial dopants or dopants in amorphous surroundings are assumed to contribute.

Further, the ratio of the backscattering signal from aligned and random RBS spectra is a measure of crystal quality. In view of RBS, a material is amorphous, when aligned and random spectrum are equal, and it is perfect crystalline, when the aligned spectrum matches the aligned spectrum of a perfect crystalline reference. This allows to calculate the relative defect concentrations in the Si lattice as a function of depth from the Si signal of the RBS spectra (see insets in Figure 1b and 1c). We used the computer code DICADA (Dechanneling In Crystals And Defect Analysis [36]) and assumed a random distribution of uncorrelated-displaced lattice atoms, which is especially valid for point defects, defect complexes and amorphous regions.

Figure 1b shows the extracted $c_{As}$ (solid lines) and $cAs_{Si}$ (dashed lines) depth profiles right after $As^+$ implantation with $N_I$ = $1x10^{17}$ $cm^{-2}$ (as implanted), and after subsequent PLM annealing at various energy densities $E_L$ from 1.0 to 2.0 $J/cm^2$, respectively. The inset in Figure 1b shows the corresponding relative defect concentrations in the Si lattice. After implantation, a Gaussian-like concentration profile with a concentration maximum of approximately $c_{As}$ ~ 18 at.% (atomic %) centered at 100 nm was found (grey line in Figure 1b). The large number of implantation-induced defects leads to complete amorphization of the first 200 nm of the Si lattice followed by a roughly 50 nm highly defective transition region (grey line in the inset of Figure 1b).

After PLM annealing, the initial As depth distribution is significantly broadened and the maximum concentration of As dopants $c_{As}$ in the near surface layer decreases with increasing laser energy density. The diffusion coefficient of As in liquid Si ($D_{As}^{L}$ ~ $10^{-4}$ $cm^2s^{-1}$ [37]) is many orders of magnitude larger in comparison to the diffusion coefficient of As in solid Si at the melting temperature, even when Si is already highly doped ($D_{As}^{S}$ ~ $10^{-15}$ to $10^{-13}$ $cm^2s^{-1}$ [38, 39]). Thus, As is dissolved into the transient melt and diffuses rapidly during PLM annealing, which broadens the initial dopant distribution over the entire melt depth, whereas diffusion of As in the solid part of Si is negligible on the time-scale of the PLM (compare $c_{As}$ profiles in Figure 1b). The melt depth, transient melt duration and velocity of the liquid-solid interface, which depend on the laser pulse length and laser energy density for a fixed laser wavelength, determine the characteristics of the reforming solid. Note, that the integral amount of As remains constant, indicating no out-diffusion or loss of As via the sample surface.

PLM annealing at $E_L$ = 1.0 J/cm$^2$ is not sufficient to melt through the entire implantation layer (melt depth: ~ 220 nm, compare inset of Figure 1b). In this case, a polycrystalline film with preferential orientation homogeneously nucleates from the undercooled melt. Although the preferential orientation is confirmed by RBS (Figure 1a), the increased backscattering yield of the aligned spectrum should not be mistaken for a high defect concentration (blue line in the inset of Figure 1b), because RBS cannot distinguish polycrystalline films from amorphous films. However, assuming that the recrystallized layer consists of grains, the amount of As on substitutional lattice sites can be roughly estimated by $cAs_{Si}/(1-n_{da})$, which yields an average value of ~8 at.% in the first 100 nm.

When the laser energy density is increased to a level high enough that the transient melt extends to the underlying single crystalline substrate ($E_L \geq$ 1.3 J/cm$^2$, melt depth ≥ 300 nm), liquid-phase epitaxial regrowth from the substrate occurs and the reforming solid is found to be a perfect single crystal (compare inset of Figure 1b). Because of the rapid cooling (quenching) process after the laser pulse, the liquid-solid interface velocity is on the order of several meter per second, thus segregation, which typically leads to a rejection of excess dopants from the liquid-solid interface into the melt, is suppressed. [40] Solute dopants are trapped at the growth front and remain dissolved in the reforming solid Si layer resulting in a super-saturation far above the equilibrium solid solubility. Especially slow diffusing dopants, such as As or P, are preferentially incorporated on substitutional lattice sites when trapped at the growth interface. [41]

A concentration of $cAs_{Si}$ ~ 8, 6, and 5 at.% of substitutional As dopants was found on average in the first 100 nm of the PLM annealed samples at $E_L$ = 1.3, 1.6 and 2.0 J/cm$^2$, respectively. The high value of 8 at.% of incorporated dopants exceeds the thermodynamic equilibrium value by a factor of 2 and is the thermodynamic equilibrium solubility for As in liquid Si. Higher dopant concentrations on Si lattice sites cannot be achieved, because the presence of a large impurity concentration may significantly reduce the epitaxial regrowth velocity. If the growth front velocity is greater than the epitaxial regrowth velocity, the atomic reordering processes are overwhelmed, and an amorphous layer is formed.

When increasing the ion fluences to 2x10$^{17}$ cm$^{-2}$ (Figure 1c), laser energy densities below 1.6 J/cm$^2$ are not sufficient to completely recrystallize the Si matrix. Here, only a laser energy density of 2.0 J/cm$^2$ lead to almost single crystalline doped Si with a small defective layer remaining at the surface (see inset in Figure 1c). Despite a rather high incorporation fraction, the maximum amount of substitutional As in the Si matrix is comparable to that by using an ion fluence of 1x10$^{17}$ cm$^{-2}$ and a laser energy density of 1.3 J/cm$^2$.

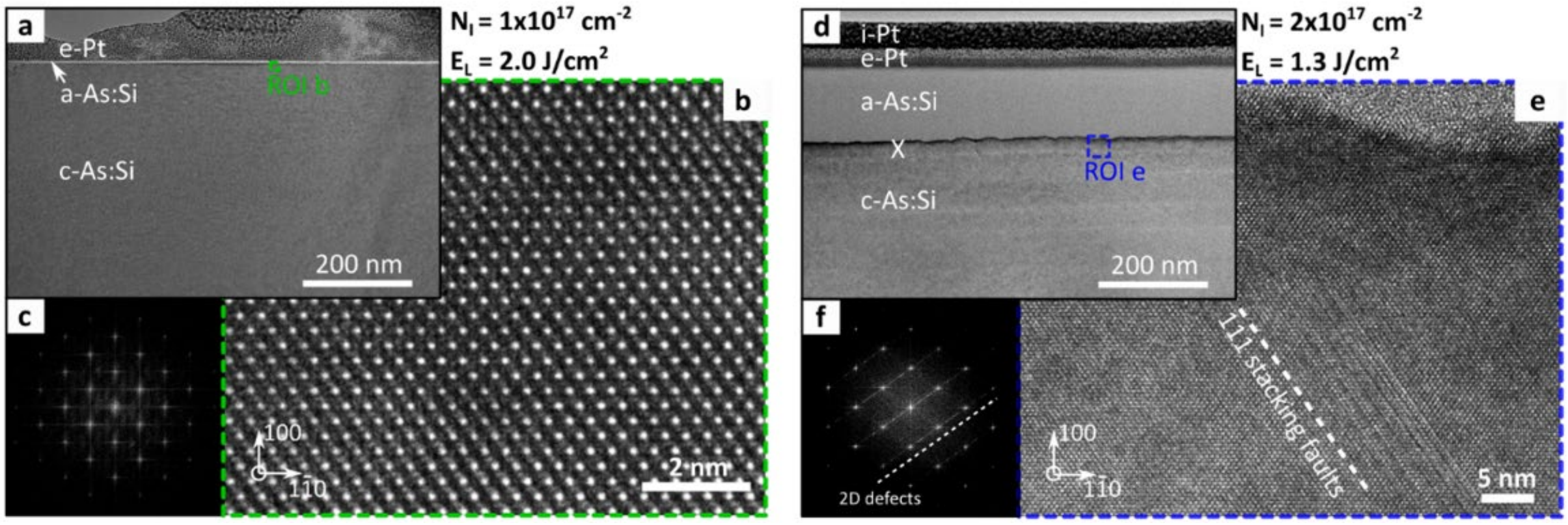

**Figure 2.** (a) Overview cross-sectional transmission electron microscopy (TEM) image of a Si sample implanted with 150 keV As$^+$ to a fluence of $N_I$ = 1x10$^{17}$ cm$^{-2}$ and subsequent PLM annealed with a laser energy density of $E_L$ = 2.0 J/cm$^2$. A thin amorphous layer (a-As:Si) on top of the fully recrystallized sample (c-As:Si) originates from surface oxidation of the sample in air. (b) High resolution TEM image of the region of interest ROI b in (a). The beam incidence is in the Si [110] direction. (c) FFT of the HRTEM image (b). (d-f) Same series of TEM analysis as (a-c) but for an ion fluence of $N_I$ = 2x10$^{17}$ cm$^{-2}$ and laser energy density of $E_L$ = 1.3 J/cm$^2$. A thick amorphous layer (a-As:Si) remains on top of the recrystallization breakdown interface (X). The prominent 2D defects at this interface are (111) stacking faults. To protect the lamella, samples were covered with e-beam deposited (e-Pt) and ion beam deposited platinum (i Pt) prior to lamella preparation.

We performed high-resolution transmission electron microscopy (HRTEM) measurements on cross-sectional lamellae taken out of two representative samples via focused ion beam (FIB) milling. **Figure 2**a shows a cross-sectional TEM overview image of a Si sample implanted with an ion fluence of $1x10^{17}$ $cm^{-2}$ and PLM annealing at 2 $J/cm^2$. The sample is perfectly homogenous and single crystalline with no defective regions across the whole implanted and PLM annealed layer. The thin amorphous layer at the surface originates from surface oxidation of the sample in air. The high-resolution TEM image close to the sample surface (Figure 2b) shows a complete recrystallized lattice meaning complete epitaxial recrystallization of highly doped Si. This is corroborated by the Fast Fourier transform (FFT) pattern in Figure 2c, which is dominated by the typical (2-fold) symmetry of Si [110].
As discussed before, for the higher ion fluence ($2x10^{17}$ $cm^{-2}$) but also smaller $E_L$ (1.3 $J/cm^2$), the dopant concentration exceeds the liquid solubility limit and, thus, the recrystallization velocity is greatly reduced. This results into the formation of an epitaxy breakdown interface between the recrystallized As:Si and a remaining amorphous As:Si layer, which can be clearly identified in Figure 2d. Further, the HRTEM image in figure 2e depicting a small region at this interface reveals extended defects in the recrystallized As:Si layer. While the FFT pattern of this image (Figure 2f) confirms such extended 2D defects, i.e., (111) stacking faults, numerous of these are detectable in the HRTEM image as lines.

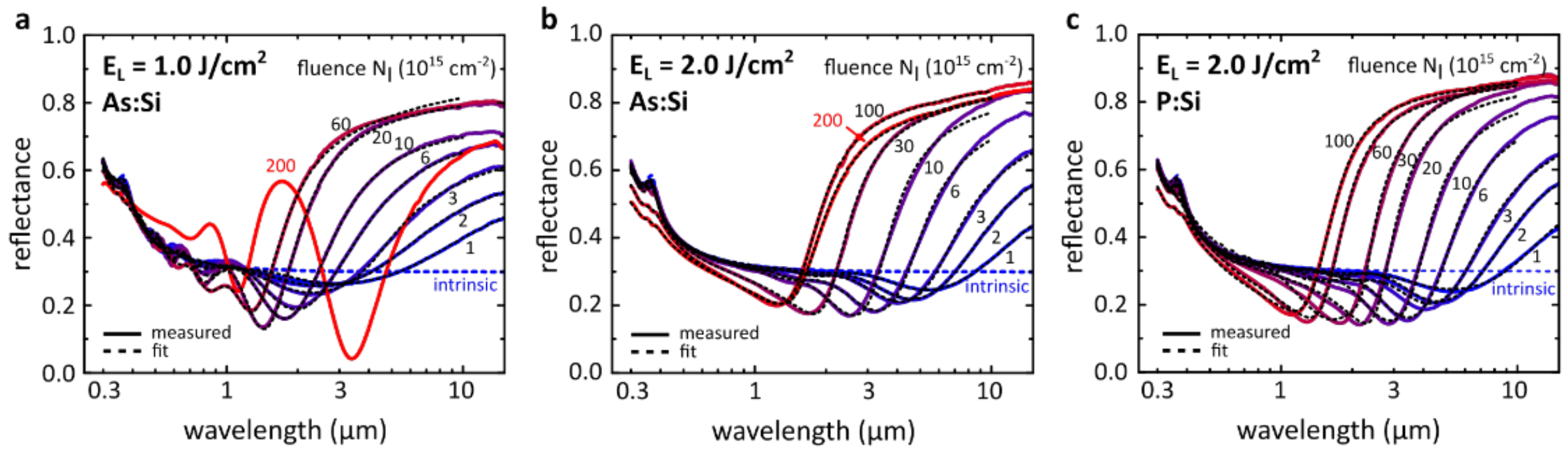

**Figure 3.** (a-c) Measured reflectance spectra and corresponding critical point (CP)-Drude model fits of Si implanted with: (a) 150 keV $As^+$ and PLM annealed at 1.0 $J/cm^2$; (b) 150 keV $As^+$ and PLM annealed at 2.0 $J/cm^2$; (c) 75 keV $P^+$ and PLM annealed at 2.0 $J/cm^2$. Reflectance spectra are given for various ion fluences ranging from $N_I = 1x10^{15}$ to $2x10^{17}$ $cm^{-2}$ as indicated, respectively. The reflectance spectrum of intrinsic Si is shown for comparison in all figures (dashed blue lines in a-c).

**Figure 3**a-b show the reflectance spectra of highly As doped Si with various ion fluences and after subsequent PLM annealing at $E_L$ = 1.0 and 2.0 $J/cm^2$, respectively. Intrinsic Si was also measured for comparison (blue dashed line). Already for low ion fluences we observe an increase in reflectance in the infrared (IR) range, which is associated with an increase of free carriers (Drude-like) within the implanted top layer of the Si substrate. With increasing ion fluence the onset of high reflectance is shifting towards shorter wavelengths, reaching the near infrared ($\lambda < 3$ µm) for ion fluences above $1x10^{16}$ $cm^{-2}$. A strong modulation of the reflectance is recognizable especially for the higher ion fluences in Figure 3a. This is caused by significant thin film interferences that occur for all samples annealed at $E_L$ = 1.0 J/cm$^2$ in the spectral range between interband transitions and the onset of high reflectance values, because the dopant distribution has a box-like profile with a sharp interface to undoped Si. Note that the strongest modulations were observed for the highest ion fluence (2x10$^{17}$ cm$^{-2}$) likely due to the remaining amorphous top layer with a sharp interface to the recrystallized doped As:Si, which is in good agreement with the cross-sectional TEM images (compare Figure 2d-e). In contrast, this interference is not occurring for samples annealed at laser energy densities of $E_L$ = 2.0 J/cm$^2$, as shown in Figure 3b. However, there are still small thin film interferences in the regime of the plasma wavelength caused by the doped Drude-like As:Si layer on top of the undoped Si substrate. The latter is corroborated by the decrease in reflectance below the initial value of intrinsic Si, followed by the strong increase of reflectance above the plasma wavelength that dominates the spectra.

Furthermore, we also prepared phosphorous implanted Si samples that were also treated with the same PLM parameters ($E_L$ = 2.0 J/cm$^2$), which is shown in figure 3c. Here, 75 keV $P^+$-ions were used in order to achieve a comparable dopant profile as for the As implantations. Although the overall shape of the reflectance curves is similar, indeed, lower onset wavelengths of the steep reflectance increase can be observed for the highest doping concentrations compared to the As case.

To further analyze the optical data of highly doped Si, and to extract the plasma wavelength and refractive index data, all reflectance spectra were fitted using the general transfer-matrix method [42] and a combination of the two-term critical-point model [43] (complex permittivity $\varepsilon_{CP}$) and Drude model ($\varepsilon_D$). A critical point model with two terms was used previously to accurately model the reflectance data of crystalline Si in the wavelength range, in which interband transitions dominate, [43, 44] whereas the Drude model accounts for the free carrier contribution, which dominates the reflectance spectra at higher wavelengths:

$$\varepsilon(\omega) = \varepsilon_{CP}(\omega) + \varepsilon_D(\omega), \qquad [1]$$

$$\varepsilon_{CP}(\omega) = \varepsilon_\infty \left[1 + \sum_k \frac{f_k(\omega_k^2 - i\gamma_k^*\omega)}{\omega_k^2 - 2i\gamma_k\omega - \omega^2}\right], \qquad [2]$$

$$\varepsilon_D(\omega) = \varepsilon_\infty \left[1 - \frac{\omega_p^2}{\omega^2 + i\Gamma\omega}\right]. \qquad [3]$$

Here, $\varepsilon_\infty$ is the background permittivity, and $\omega_p$ is the screened plasma frequency, which is proportional to the free-carrier concentration $N$: $\omega_p^2 = Ne^2/(\varepsilon_0\varepsilon_\infty m^*)$, where $e$ is the elementary charge, $\varepsilon_0$ is the vacuum permittivity, and $m^*$ is the effective mass, which was assumed to be doping independent and set to 0.27 times the free electron mass. [13] $\Gamma$ is a damping factor associated with the mobility of the charge carriers $\mu$: $\Gamma \propto \mu^{-1}$. In the critical point model $\omega_k$, $f_k$, $\gamma_k$*, and $\gamma_k$ is the frequency, oscillator strength, induced polarization, and electronic damping of the $k$-th critical point, associated with interband transitions, respectively. To reduce the number of variables, the frequencies of both critical points were fixed to values found for intrinsic Si $\omega_1 = 3.37$ eV and $\omega_2 = 4.24$ eV, respectively.

A model consisting of a uniformly doped 100 nm surface layer, a gradient-index transition layer with variable thickness, and an undoped semi-infinite Si substrate was used to fit the reflectance data. The results did not significantly change by adding additional layers or dividing the surface layer. Especially at wavelengths, at which the reflectance is high, only the surface layer contributes to the reflectance spectrum and the transition layer can be completely neglected due to the high losses in the metal-like doped layer. However, the gradient-index transition layer between the uniformly doped top layer and the undoped substrate is necessary to model the disappearance of thin film interference effects in the visible spectral region for samples PLM annealed at high laser energy densities.

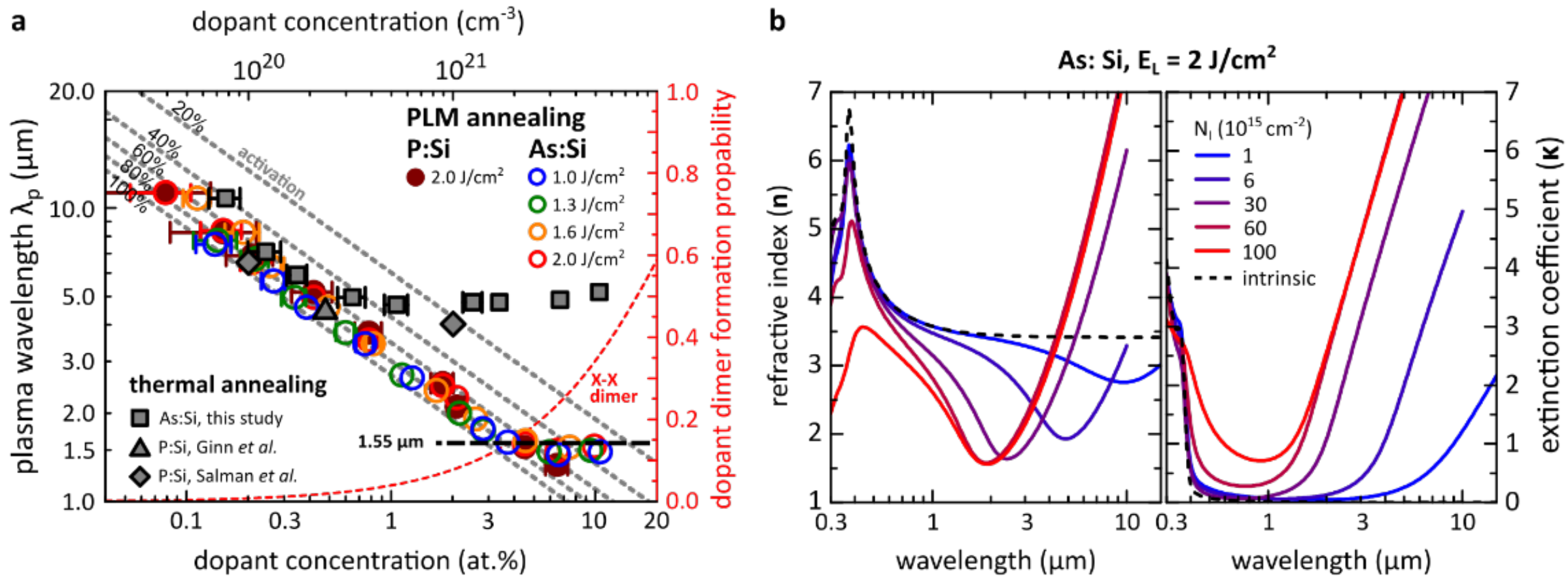

**Figure 4.** (a) Summary of plasma wavelengths $\lambda_p$ extracted from CP-Drude model fits of all reflectance data acquired for this study, respectively: 150 keV As:Si PLM annealed at $E_L$ between 1 and 2 J/cm$^2$ (open circles); 75 keV P:Si PLM annealed at $E_L = 2$ J/cm$^2$ (filled circles); 150 keV As:Si oven annealed at 900°C for 1h under high vacuum (filled squares). For comparison, plasma wavelengths of highly doped Si experimentally determined by Ginn et al. (filled triangle, [45]) and Salman et al. (filled rhombs, [13]) are included. Grey dashed lines indicate plasma wavelengths that correspond to a certain dopant activation fraction. The red dashed line indicates the probability of dopant atoms to share nearest neighbor sites (dopant dimer formation, X-X) calculated for a random distribution of dopants on Si lattice sites. (b) Refractive index $n$ and extinction coefficient $\kappa$ of intrinsic Si (dashed lines) compared to $n$ and $\kappa$ of highly doped Si implanted with 150 keV As$^+$ and PLM annealed at $E_L = 2.0$ J/cm$^2$ for various ion fluences $N_I$, as indicated.

In general, we find excellent agreement between our fits and experimental data. **Figure 4**a summarizes all extracted plasma wavelengths as a function of the total amount of As dopants $c_{As}$. Large error bars at low dopant concentrations reflect the limits of RBS measurements.

Energy dispersive X-ray analysis (EDX) in the TEM on As and P doped samples reveal a similar dopant profile for comparable implantation and PLM annealing conditions (for more details see supporting information). Therefore, the P dopant concentration was estimated as the corresponding As doping concentration. To account for the uncertainty of this estimation, the error bars were chosen accordingly. For comparison, dashed grey lines in Figure 4a give the calculated plasma wavelength assuming 100, 80, 60, 40, and 20% dopant activation, respectively. Up to a dopant concentration of 4 at.% most of the samples show a dopant activation of ≥ 80%. The lowest plasma wavelength of 1430 nm was indeed achieved for phosphorus doped Si with an ion fluence of $1x10^{17}$ $cm^{-2}$ and laser energy density of 2 $J/cm^2$. For dopant concentrations above 4 at.% the activation quickly decreases and no further decrease of the plasma wavelength (increase in free carrier concentration) can be achieved. Most likely, the reason is random dimer formation that induce localized deep-level states in the band gap, compensating the free carriers. [46] Thus, we calculated the probability for random occupation of the Si lattice (right, red axis in Figure 4a). For high dopant concentrations, the probability increases that by chance the first neighbor of an impurity in the lattice is an impurity. It was shown that especially the formation of negatively charged clusters comprising two As atoms saturates/decreases the free electron density. [46] Thus, only isolated impurities contribute to the free carrier concentration. [17]
For comparison we included the extracted plasma wavelength of As doped samples that were long-term annealed in a furnace at 900°C for 1h under high vacuum (<4 x $10^{-4}$ mbar). Here, equilibrium concentrations cannot be reached within 1h, therefore, the dopant concentration significantly exceeds the equilibrium solid solubility of As in Si. However, we find a minimum plasma wavelength of ~ 5 µm for all dopant concentrations above 0.5 at.%. This can be attributed to diffusion of As and their cluster formation. We also included literature data of oven annealed n-type doped Si. [13, 45] Although a slightly lower plasma frequency with higher dopant activation was achieved, the overall achievable plasma wavelength was found to be ~ 4 µm.
Furthermore, we extracted the complex refractive index, i.e., refractive index $n$ and extinction coefficient $\kappa$, for intrinsic Si and As:Si with various ion fluences and subsequent PLM annealed with $E_L$ = 2.0 $J/cm^2$, which is shown in Figure 4b. While the strong Drude-like contribution of free carries can clearly be identified in the infrared region above the respective plasma wavelengths, the contribution of interband transitions is mostly unchanged. Only for the highest ion fluences ($6x10^{16}$ and $1x10^{17}$ $cm^{-2}$) a significant decrease of the real part of the refractive index in the visible spectral region is apparent due to a remaining amorphous layer as previously discussed.

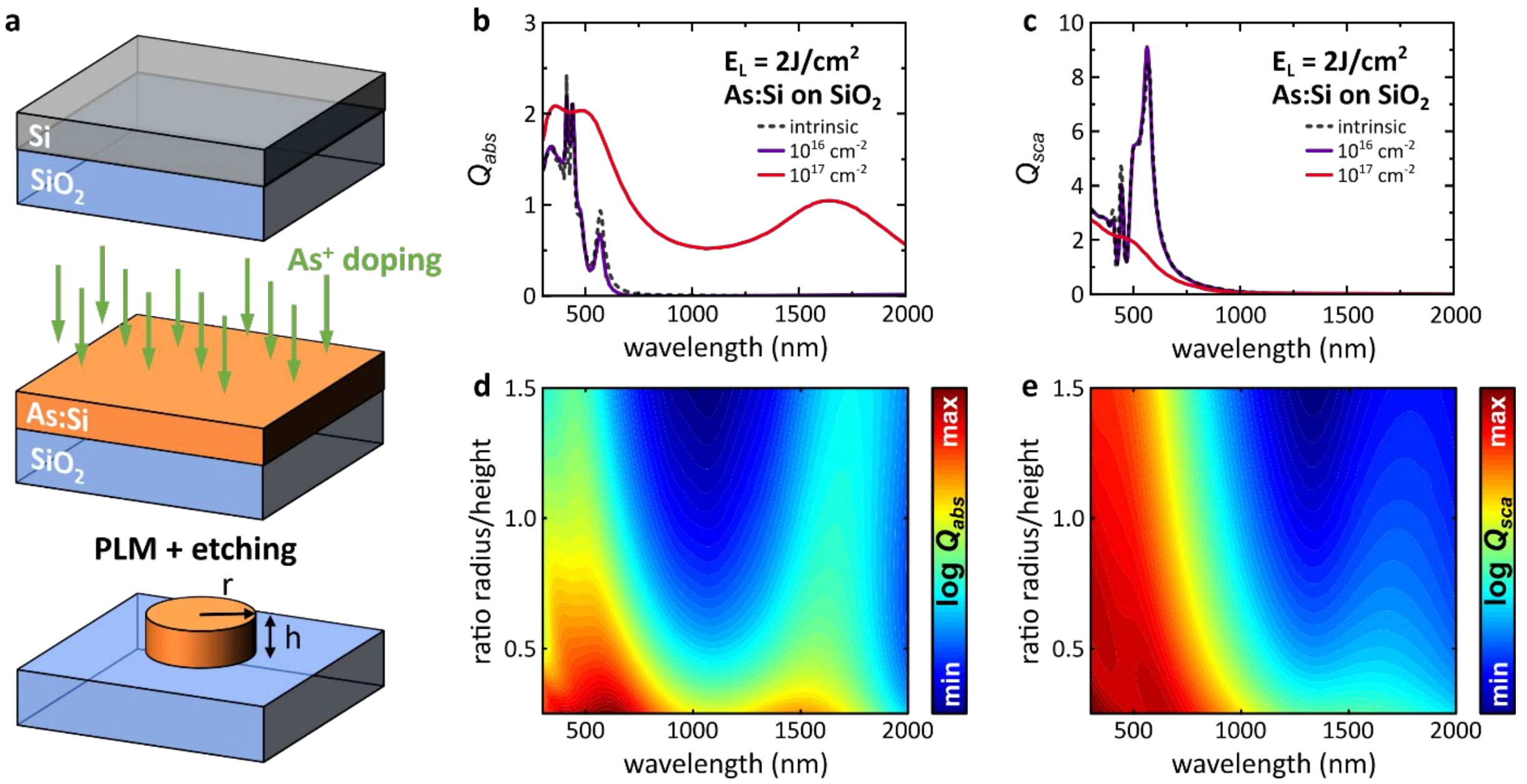

**Figure 5.** (a) Schematic drawing for a possible fabrication process to achieve As-doped Si nanoparticles. (b) Absorption and (c) scattering efficiency spectra for a cylinder (radius $r$ = 70 nm and height $h$ = 100 nm) located on a $SiO_2$ substrate for intrinsic Si (black dashed line) and Si implanted with 150 keV $As^+$ to a fluence of $N_I$ = 1x10$^{16}$ cm$^{-2}$ (purple line) and $N_I$ = 1x10$^{17}$ cm$^{-2}$ (red line) at a laser energy density $E_L$ of 2.0 J/cm$^2$, respectively. (d) Logarithmic scaled absorption and (e) scattering efficiency spectra for Si implanted with 150 keV $As^+$ to a fluence of $N_I$ = 1x10$^{17}$ cm$^{-2}$ at a laser energy density $E_L$ of 2.0 J/cm$^2$, as a function of the aspect ratio of the cylinder (radius/height).

The extracted dielectric functions were used to demonstrate the potential of our approach to achieve highly-doped As:Si nanoparticles. Here, we propose a fabrication process as shown in **Figure 5**a: first SOI (silicon-on-insulator, Si on $SiO_2$) substrates are irradiated with As resulting in an amorphous As:Si layer on top of the $SiO_2$ layer. Next, local PLM annealing using a focused laser beam can be used to crystallize pre-defined, restricted regions within the amorphous host. [47] In this way arbitrary shapes and geometries can be directly written into the doped Si, but being diffraction limited by the used annealing laser. By taking advantage of the selective etching properties of amorphous and crystalline Si, [48, 49] only the PLM treated regions remain while the amorphous surrounding is removed. Thus, we can use this method to create plasmonic nanostructures based on highly doped Si while no additional masking step is necessary.

We performed finite-difference time-domain (FDTD) simulations using the commercial software package Lumerical to demonstrate the potential of our approach for near-infrared plasmonics. Figure 5b and c represent the absorption and scattering efficiency spectra for a cylinder located on a $SiO_2$ substrate for intrinsic Si (black dashed line) and Si implanted with 150 keV $As^+$ to an ion fluence of $N_I = 1x10^{16}$ $cm^{-2}$ (purple line) and $1x10^{17}$ $cm^{-2}$ (red line) at a laser energy density $E_L$ of 2.0 J/$cm^2$, respectively. The height and radius of the cylinder correspond to $h$ = 100 nm and $r$ = 70 nm. The structure is illuminated with a plane wave under normal incidence, i.e., propagating along the axis of the cylinder, from the air to the $SiO_2$ substrate, and linearly polarized and perpendicularly to the axis of the cylinder. The spectral response of intrinsic Si and $As^+$ lower doped Si (with $N_I = 1x10^{16}$ $cm^{-2}$) is found to be the same in both cases. Here, the sharp resonances around 500 nm are Mie resonances, corresponding to electric and magnetic dipolar modes. However, we achieve a distinct electromagnetic response for the case of highly doped As:Si implanted to a fluence of $N_I$ = $1x10^{17}$ $cm^{-2}$. In addition to a broader Mie resonance corresponding to a dipolar electric mode, the most notable feature is the appearance of a resonance in the near-infrared region ($\lambda \approx 1550$ nm) in the absorption efficiency spectra. This suggests that it is possible to attain a plasmonic response in the near-infrared range by means of hyper-doped Si. The absorption and scattering efficiency spectra for Si implanted with 150 keV $As^+$ to a fluence of $N_I = 1x10^{17}$ $cm^{-2}$ at a laser energy density $E_L$ of 2.0 J/$cm^2$ is shown in Figure 5d and e as a function of the aspect ratio of the cylinder (radius/height) and on a logarithmic scale. A spectral shift of the plasmonic resonance towards larger wavelengths with increasing aspect ratio is observed in the absorption spectra. Note that on a logarithmic scale, the scattering spectra show also a weak plasmonic response (see Figure 5e) that is not visible in the linearly scaled scattering spectra of Figure 5c. For most conventionally used semiconductor materials (Si or other semiconductor compounds such as GaAs, AlSb and GaP), absorption is negligible in the near-infrared region. For other materials like Ge, the absorption is negligible above $\lambda \approx 1500$ nm. This makes that the extinction efficiency, which is the sum of the scattering and absorption efficiencies ($Q_{ext} = Q_{abs}+Q_{sca}$), is given by the scattering efficiency at near-infrared wavelengths, where the absorption is negligible for intrinsic high-refractive index dielectric nanoparticles. Therefore, introducing a resonance in the absorption spectrum can be used to enhance the efficiency of photodetectors and achieve perfect absorption at telecommunication wavelengths.

## 3. Conclusion

In this study, we systematically investigated the highest possible concentration of n-type dopants that can be incorporated into the silicon (Si) lattice using a combination of ion implantation and pulsed laser melting (PLM) annealing processes. We examined the depth distribution of arsenic (As) dopants using random and channeling Rutherford backscattering (RBS) spectroscopy and determined the total amount of dopants in the Si lattice and the amount located on substitutional lattice sites. During PLM annealing, we observed broadening of the dopant distribution and a decrease in the concentration of As dopants near the surface as the laser energy density increased. The annealing process facilitated the dissolution and rapid diffusion of As in the transient melt, while diffusion in the solid part of Si was negligible. The characteristics of the solid reforming were determined by factors such as the melt depth, transient melt duration, and velocity of the solidification front. At lower laser energy densities, partial recrystallization resulted in a polycrystalline film with preferential orientation. However, at higher laser energy densities, complete epitaxial regrowth from the substrate led to the formation of a single crystal. Notably, slow diffusing dopants such as As and P were preferentially incorporated on substitutional lattice sites during the regrowth process. TEM measurements confirmed the complete epitaxial recrystallization of highly doped Si, while defects and extended defects were observed in samples with high ion fluences and smaller laser energy densities. Our analysis of the reflectance spectra revealed a large increase in free carrier concentration in the doped Si samples. We extracted the plasma wavelength, which correlates with the free carrier concentration, from the reflectance data. Our results show an achievable plasma wavelength of approximately 1.5 μm for dopant concentrations above 4 at.%. Furthermore, we determined the complex refractive index of the doped Si samples. Based on these findings, we demonstrated the potential for fabricating hyper-doped Si nanoparticles using our proposed process. By selectively crystallizing regions within an amorphous host and removing the surrounding amorphous material, we could create plasmonic nanostructures without the need for additional masking steps. Using finite-difference time-domain (FDTD) simulations, we illustrated the potential of our approach for near-infrared plasmonics. The simulations showed distinct electromagnetic responses and plasmonic resonances in the near-infrared range for hyper-doped Si nanoparticles. In conclusion, we demonstrate the enormous potential of our approach in designing CMOS compatible, plasmonic nanostructures operating at telecommunication wavelengths.

## 4. Experimental Section

Intrinsic (001) Si samples (FZ, > 105 Ωcm, single-side-polished, 1x1 cm$^2$) were implanted with either 150 keV $As^+$ or 75 keV $P^+$ ions to high ion fluences ranging from $N_I = 1\text{x}10^{15}$ to $2\text{x}10^{17}$ cm$^{-2}$, respectively. The ion energies were chosen by means of Monte-Carlo simulations using the software package SRIM (Stopping and Range of Ions in Matter) [50] to ensure comparable doping profiles with a concentration maximum in a depth of 100 nm. All implantations were performed at room temperature under an angle of 7° to prevent possible channeling effects. The beam current density was kept at ~ 1 μA/cm$^2$ to avoid intense beam-heating of the samples, which would result in dynamic damage annealing.

After implantation, most of the samples were annealed using pulsed laser melting to recover the crystal lattice and to activate the dopants. PLM annealing was achieved with several 15 ns long pulses of an Yb:YAG diode-pumped solid-state laser (343 nm, 10 kHz) and various energy densities between $E_L$ = 1.0 and 2.0 J/cm$^2$. To achieve homogeneous annealing, a INNOVAVENT VOLCANO® laser optics systems was used to create a Gaussian line beam of ~ 20 μm FWHM and a length of ~ 20 mm. The samples were moved at a constant speed of 20 mm/s through the line beam with the sample surface in the focal plane, which results in a pulse overlap of roughly 90% (~10 pulses per area). For comparison, some of the implanted samples were oven annealed at 900°C for 1 h under high vacuum (<4x10$^{-4}$ mbar) to avoid oxidization.

The structure and elemental composition of all samples, both after implantation and after annealing, were characterized by means of Rutherford backscattering spectroscopy (RBS) and channeling measurements using a 1.4 MeV $^4He^+$ ion beam. The depth distribution and total amount of dopants in the Si samples were extracted from RBS spectra taken in random direction (3° off axis, sample rotated during measurements). The number of substitutional dopants on Si lattice sides and the damage fraction were determined from aligned spectra, which were measured by carefully aligning the (001) axis of the Si substrate with the incident $^4He^+$ ion beam.

Additionally, the structure, morphology and elemental composition of selected samples were evaluated after PLM annealing by means of cross-sectional transmission electron microscopy (TEM) using a JEOL NEOARM 200F TEM equipped with dual EDX (energy dispersive X-ray spectroscopy) detectors for elemental analysis. The cross-sectional TEM lamellae were prepared using a focused ion beam system FEI Helios Nanolab 600i. The lamellae were cut out of the Si samples parallel to the Si <110> direction.

For optical characterization of all samples both after implantation and after annealing, reflectance spectra were recorded at near normal incidence (12°) in the spectral range from 300 nm to 15 μm using a Varian Cary 5000 UV-VIS spectrometer and a Varian 640 FT-IR spectrometer.

*Simulation details:*

The cylinder was located in the center of the three-dimensional FDTD simulation domain on a $SiO_2$ substrate. The optical properties of the cylinder were obtained through the experimental measurements. A total-field scattered-field (TFSF) source was used for the illumination. A plane wave impinged from the air region to the substrate direction at normal incidence. The incident radiation was linearly polarized perpendicularly to the cylinder axis. The scattering efficiency (defined as the scattering cross-section $\sigma_{sca}$ divided by the geometrical cross-section $\sigma_{geo}$, $Q_{sca} = \sigma_{sca}/\sigma_{geo}$ cross-section) was obtained by means of a six-monitor box surrounding the source. The scattered power was calculated via Poynting vector integration of the scattered field over the monitors. The ratio of the scattered power to the incident intensity yielded the scattering cross-section. The absorption efficiency was obtained in the same way as the scattering efficiency with the only difference that the six-monitor box was inside the source, i.e., between the source and the cylinder. Perfectly matched layers (PMLs) were employed to absorb the scattered field. An automatic non-uniform mesh refinement (5 nm) was used for the region surrounding the cylinder to attain the convergence of the solution.

**Supporting Information**

Supporting Information is available from the Wiley Online Library or from the author.

**Acknowledgements**

This work was financed by the German Federal Ministry of Education and Research within the funding program Photonics Research Germany 13N14922 and the Deutsche Forschungsgemeinschaft (DFG) within the frame of the international graduate school IRTG 2675 “Meta-Active”. A. Barreda gratefully acknowledges financial support from Spanish national project No. PID2022-137857NA-I00. A. Barreda thanks MICINN for the Ramon y Cajal Fellowship (grant No. RYC2021-030880-I). J. Rensberg and A. Barreda contributed equally to this work.

Received: ((will be filled in by the editorial staff))
Revised: ((will be filled in by the editorial staff))
Published online: ((will be filled in by the editorial staff))

## References

1 Lal, S., Link, S., Halas, N.J. (2007) Nano-optics from sensing to waveguiding. *Nature Photon*, **1** (11), 641–648.

2 Brongersma, M.L. and Shalaev, V.M. (2010) Applied physics. The case for plasmonics. *Science (New York, N.Y.)*, **328** (5977), 440–441.

3 Ozbay, E. (2006) Plasmonics: merging photonics and electronics at nanoscale dimensions. *Science (New York, N.Y.)*, **311** (5758), 189–193.

4 Schuller, J.A., Barnard, E.S., Cai, W., Jun, Y.C., White, J.S., Brongersma, M.L. (2010) Plasmonics for extreme light concentration and manipulation. *Nature materials*, **9** (3), 193–204.

5 Leonhardt, U. (2006) Optical conformal mapping. *Science (New York, N.Y.)*, **312** (5781), 1777–1780.

6 West, P.R., Ishii, S., Naik, G.V., Emani, N.K., Shalaev, V.M., Boltasseva, A. (2010) Searching for better plasmonic materials. *Laser & Photonics Reviews*, **4** (6), 795–808.

7 Naik, G.V., Shalaev, V.M., Boltasseva, A. (2013) Alternative plasmonic materials: beyond gold and silver. *Advanced materials (Deerfield Beach, Fla.)*, **25** (24), 3264–3294.

8 Soref, R. (2006) The Past, Present, and Future of Silicon Photonics. *IEEE J. Select. Topics Quantum Electron.*, **12** (6), 1678–1687.

9 Barber, H.D. (1967) Effective mass and intrinsic concentration in silicon. *Solid-State Electronics*, **10** (11), 1039–1051.

10 Hryciw, A., Jun, Y.C., Brongersma, M.L. (2010) Plasmonics: Electrifying plasmonics on silicon. *Nature materials*, **9** (1), 3–4.

11 Dionne, J.A., Sweatlock, L.A., Sheldon, M.T., Alivisatos, A.P., Atwater, H.A. (2010) Silicon-Based Plasmonics for On-Chip Photonics. *IEEE J. Select. Topics Quantum Electron.*, **16** (1), 295–306.

12 Soref, R. (2010) Mid-infrared photonics in silicon and germanium. *Nature Photon*, **4** (8), 495–497.

13 Salman, J., Hafermann, M., Rensberg, J., Wan, C., Wambold, R., Gundlach, B.S., Ronning, C., Kats, M.A. (2018) Flat Optical and Plasmonic Devices Using Area-Selective Ion-Beam Doping of Silicon. *Advanced Optical Materials*, **6** (5).

14 Cleary, J.W., Peale, R.E., Shelton, D.J., Boreman, G.D., Smith, C.W., Ishigami, M., Soref, R., Drehman, A., Buchwald, W.R. (2010) IR permittivities for silicides and doped silicon. *J. Opt. Soc. Am. B*, **27** (4), 730.

15 Trumbore, F.A. (1960) Solid Solubilities of Impurity Elements in Germanium and Silicon*. *Bell System Technical Journal*, **39** (1), 205–233.

16 Williams, J.S. and Elliman, R.G. (1981) Limits to solid solubility in ion implanted silicon. *Nuclear Instruments and Methods*, **182-183**, 389–395.

17 Suzuki, K., Tada, Y., Kataoka, Y., Kawamura, K., Nagayama, T., Nagayama, S., Magee, C.W., Buyuklimanli, T.H., Mueller, D.C., Fichtner, W., Zechner, C. (2007) Maximum Active Concentration of Ion-Implanted Phosphorus During Solid-Phase Epitaxial Recrystallization. *IEEE Trans. Electron Devices*, **54** (8), 1985–1993.

18 Lietoila, A., Gibbons, J.F., Sigmon, T.W. (1980) The solid solubility and thermal behavior of metastable concentrations of As in Si. *Applied Physics Letters*, **36** (9), 765–768.

19 Kooi, E. (1964) Formation and Composition of Surface Layers and Solubility Limits of Phosphorus During Diffusion in Silicon. *J. Electrochem. Soc.*, **111** (12), 1383.

20 Maekawa, S. (1962) Diffusion of Phosphorus into Silicon. *J. Phys. Soc. Jpn.*, **17** (10), 1592–1597.

21 Zhou, S., Pi, X., Ni, Z., Luan, Q., Jiang, Y., Jin, C., Nozaki, T., Yang, D. (2015) Boron- and Phosphorus-Hyperdoped Silicon Nanocrystals. *Part. Part. Syst. Charact.*, **32** (2), 213–221.

22 Rowe, D.J., Jeong, J.S., Mkhoyan, K.A., Kortshagen, U.R. (2013) Phosphorus-doped silicon nanocrystals exhibiting mid-infrared localized surface plasmon resonance. *Nano letters*, **13** (3), 1317–1322.

23 Poumirol, J.-M., Majorel, C., Chery, N., Girard, C., Wiecha, P.R., Mallet, N., Monflier, R., Larrieu, G., Cristiano, F., Royet, A.-S., Alba, P.A., Kerdiles, S., Paillard, V., Bonafos, C. (2021) Hyper-Doped Silicon Nanoantennas and Metasurfaces for Tunable Infrared Plasmonics. *ACS Photonics*, **8** (5), 1393–1399.

24 Liu, F., Prucnal, S., Berencén, Y., Zhang, Z., Yuan, Y., Liu, Y., Heller, R., Böttger, R., Rebohle, L., Skorupa, W., Helm, M., Zhou, S. (2017) Realizing the insulator-to-metal transition in Se-hyperdoped Si via non-equilibrium material processing. *J. Phys. D: Appl. Phys.*, **50** (41), 415102.

25 Wang, M., Debernardi, A., Berencén, Y., Heller, R., Xu, C., Yuan, Y., Xie, Y., Böttger, R., Rebohle, L., Skorupa, W., Helm, M., Prucnal, S., Zhou, S. (2019) Breaking the Doping Limit in Silicon by Deep Impurities. *Phys. Rev. Applied*, **11** (5).

26 Sánchez, K., Aguilera, I., Palacios, P., Wahnón, P. (2010) Formation of a reliable intermediate band in Si heavily coimplanted with chalcogens (S, Se, Te) and group III elements (B, Al). *Phys. Rev. B*, **82** (16).

27 Olea, J., González-Díaz, G., Pastor, D., Mártil, I. (2009) Electronic transport properties of Ti-impurity band in Si. *J. Phys. D: Appl. Phys.*, **42** (8), 85110.

28 García-Hemme, E., García-Hernansanz, R., Olea, J., Pastor, D., Del Prado, A., Mártil, I., González-Díaz, G. (2013) Far infrared photoconductivity in a silicon based material: Vanadium supersaturated silicon. *Applied Physics Letters*, **103** (3).
29 Komarov, F.F., Nechaev, N.S., Ivlev, G.D., Vlasukova, L.A., Parkhomenko, I.N., Wendler, E., Romanov, I.A., Berencén, Y., Pilko, V.V., Zhigulin, D.V., Komarov, A.F. (2020) Structural and optical properties of Si hyperdoped with Te by ion implantation and pulsed laser annealing. *Vacuum*, **178**, 109434.
30 Simmons, C.B., Akey, A.J., Krich, J.J., Sullivan, J.T., Recht, D., Aziz, M.J., Buonassisi, T. (2013) Deactivation of metastable single-crystal silicon hyperdoped with sulfur. *Journal of Applied Physics*, **114** (24).
31 Wang, K.-F., Shao, H., Liu, K., Qu, S., Wang, Y., Wang, Z. (2015) Possible atomic structures responsible for the sub-bandgap absorption of chalcogen-hyperdoped silicon. *Applied Physics Letters*, **107** (11).
32 Yang, W., Ferdous, N., Simpson, P.J., Gaudet, J.M., Hudspeth, Q., Chow, P.K., Warrender, J.M., Akey, A.J., Aziz, M.J., Ertekin, E., Williams, J.S. (2019) Evidence for vacancy trapping in Au-hyperdoped Si following pulsed laser melting. *APL Materials*, **7** (10).
33 Wang, M., Hübner, R., Xu, C., Xie, Y., Berencén, Y., Heller, R., Rebohle, L., Helm, M., Prucnal, S., Zhou, S. (2019) Thermal stability of Te-hyperdoped Si: Atomic-scale correlation of the structural, electrical, and optical properties. *Phys. Rev. Materials*, **3** (4).
34 White, C.W., Pronko, P.P., Wilson, S.R., Appleton, B.R., Narayan, J., Young, R.T. (1979) Effects of pulsed ruby-laser annealing on As and Sb implanted silicon. *Journal of Applied Physics*, **50** (5), 3261–3273.
35 Jeynes, C., Barradas, N.P., Marriott, P.K., Boudreault, G., Jenkin, M., Wendler, E., Webb, R.P. (2003) Elemental thin film depth profiles by ion beam analysis using simulated annealing - a new tool. *J. Phys. D: Appl. Phys.*, **36** (7), R97-R126.
36 Gärtner, K. (2005) Modified master equation approach of axial dechanneling in perfect compound crystals. *Nuclear Instruments and Methods in Physics Research Section B: Beam Interactions with Materials and Atoms*, **227** (4), 522–530.
37 Garandet, J.P. (2007) New Determinations of Diffusion Coefficients for Various Dopants in Liquid Silicon. *Int J Thermophys*, **28** (4), 1285–1303.
38 Fair, R.B. and Weber, G.R. (1973) Effect of complex formation on diffusion of arsenic in silicon. *Journal of Applied Physics*, **44** (1), 273–279.
39 Swaminathan, B., Saraswat, K.C., Dutton, R.W., Kamins, T.I. (1982) Diffusion of arsenic in polycrystalline silicon. *Applied Physics Letters*, **40** (9), 795–798.
40 Wood, R.F. (1982) Macroscopic theory of pulsed-laser annealing. III. Nonequilibrium segregation effects. *Phys. Rev. B*, **25** (4), 2786–2811.
41 Claeys, C. and Simoen, E. (2018) *Metal Impurities in Silicon- and Germanium-Based Technologies*, Springer International Publishing, Cham.
42 Katsidis, C.C. and Siapkas, D.I. (2002) General transfer-matrix method for optical multilayer systems with coherent, partially coherent, and incoherent interference. *Applied optics*, **41** (19), 3978–3987.
43 Deinega, A. and John, S. (2012) Effective optical response of silicon to sunlight in the finite-difference time-domain method. *Optics letters*, **37** (1), 112–114.
44 Leng, J., Opsal, J., Chu, H., Senko, M., Aspnes, D.E. (1998) Analytic representations of the dielectric functions of materials for device and structural modeling. *Thin Solid Films*, **313-314**, 132–136.
45 Ginn, J.C., Jarecki, R.L., Shaner, E.A., Davids, P.S. (2011) Infrared plasmons on heavily-doped silicon. *Journal of Applied Physics*, **110** (4).
46 Komarov, F.F., Velichko, O.I., Dobrushkin, V.A., Mironov, A.M. (2006) Mechanisms of arsenic clustering in silicon. *Phys. Rev. B*, **74** (3).
47 Florian, C., Fischer, D., Freiberg, K., Duwe, M., Sahre, M., Schneider, S., Hertwig, A., Krüger, J., Rettenmayr, M., Beck, U., Undisz, A., Bonse, J. (2021) Single Femtosecond Laser-Pulse-Induced Superficial Amorphization and Re-Crystallization of Silicon. *Materials (Basel, Switzerland)*, **14** (7).
48 Agrawal, N., Tarey, R.D., Chopra, K.L. (1991) Etching of crystalline and amorphous silicon in CClF3 plasma. *Thin Solid Films*, **198** (1-2), 393–399.
49 Martinez-Jimenez, G., Franz, Y., Runge, A.F.J., Ceschia, M., Healy, N., Oo, S.Z., Tarazona, A., Chong, H.M.H., Peacock, A.C., Mailis, S. (2019) Photonic micro-structures produced by selective etching of laser-crystallized amorphous silicon. *Opt. Mater. Express*, **9** (6), 2573.
50 Ziegler, J.F., Ziegler, M.D., Biersack, J.P. (2010) SRIM – The stopping and range of ions in matter (2010). *Nuclear Instruments and Methods in Physics Research Section B: Beam Interactions with Materials and Atoms*, **268** (11-12), 1818–1823.

Hyper-doping of silicon is fabricated by a combination of ion implantation and pulsed laser melting annealing processes. By this means we achieve a plasma wavelength of 1.5 µm, which is promising for near-infrared plasmonic applications. We propose a fabrication process for CMOS-compatible, hyper-doped silicon nanoparticles for advanced photonic and optoelectronic applications.

J. Rensberg, A. Barreda, K. Wolf, A. Undiz, J. Salfeld, S. Geburt, I. Staude, C. Ronning, and M. Hafermann *

**Hyper-doping of Silicon for Plasmonics in the Telecommunication Range**

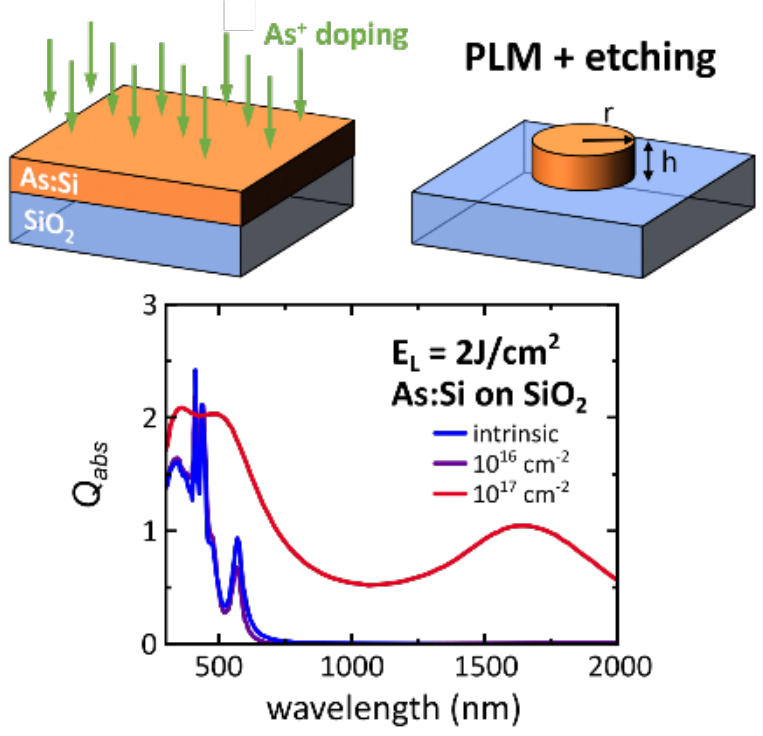

# Supporting Information

## Hyper-doping of Silicon for Plasmonics in the Telecommunication Range

*Jura Rensberg, Angela Barreda, Kevin Wolf, Andreas Undisz, Jürgen Salfeld, Sebastian Geburt, Isabelle Staude, Carsten Ronning, and Martin Hafermann**

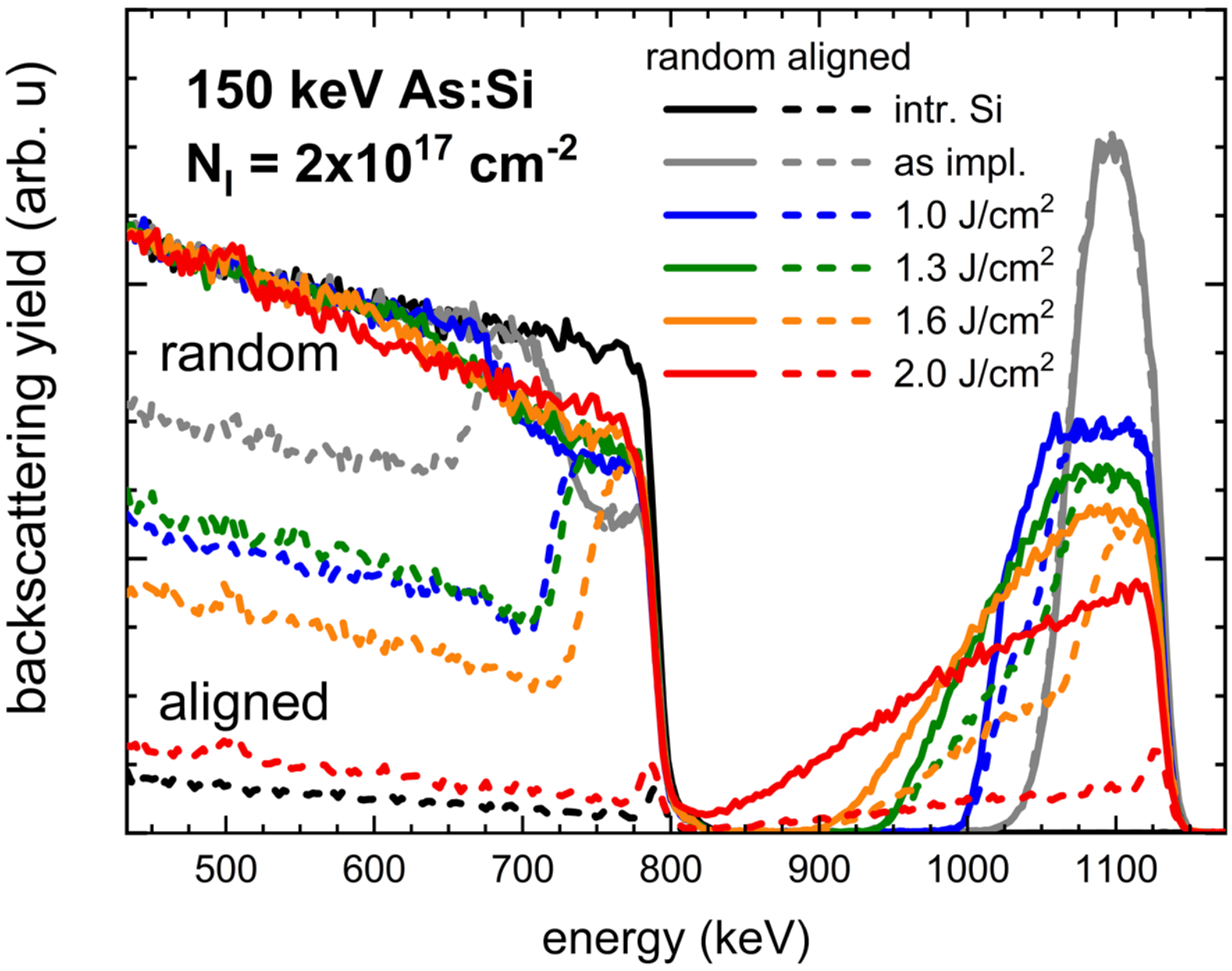

**Figure S1.** Random and aligned RBS spectra of intrinsic Si compared to random and aligned RBS spectra of Si implanted with 150 keV $As^+$ to a fluence of $N_I$ = 2x10$^{17}$ cm$^{-2}$ before (as implanted) and after subsequent pulse laser melting (PLM) annealing with different laser energy densities from 1.0 to 2.0 J/cm$^2$.